\documentclass[aps,prl,twocolumn,groupedaddress,amssymb]{revtex4}
\usepackage{latexsym}
\usepackage{amsbsy}
\usepackage{amsmath}
\usepackage{graphicx}
\usepackage{color}

\usepackage{color}

\begin{document}
\date{\today}

\title{Skyrmions at the edge: Confinement effects in Fe/Ir(111)}

\author{Julian Hagemeister}
\author{Davide Iaia}
\altaffiliation{Department of Physics, University of Illinois, 104 South Goodwin Avenue, Urbana, IL 61801, USA}
\author{Elena~Y.~Vedmedenko}
\author{Kirsten~von~Bergmann}
\author{Andr\'{e}~Kubetzka}
\email{kubetzka@physnet.uni-hamburg.de}
\author{Roland Wiesendanger}

\affiliation{Department of Physics, University of Hamburg, D-20355 Hamburg, Germany}

\date{\today}

\begin{abstract}
We have employed spin-polarized scanning tunneling microscopy and Monte-Carlo simulations to investigate the effect of lateral confinement onto the nanoskyrmion lattice in Fe/Ir(111). We find a strong coupling of one diagonal of the square magnetic unit cell to the close-packed edges of Fe nanostructures. In triangular islands this coupling in combination with the mismatching symmetries of the islands and of the square nanoskyrmion lattice leads to frustration and triple-domain states. In direct vicinity to ferromagnetic NiFe islands, the surrounding skyrmion lattice forms additional domains. In this case a side of the square magnetic unit cell prefers a parallel orientation to the ferromagnetic edge. These experimental findings can be reproduced and explained by Monte-Carlo simulations. Here, the single-domain state of a triangular island is lower in energy, but nevertheless multi-domain states occur due to the combined effect of entropy and an intrinsic domain wall pinning arising from the skyrmionic character of the spin texture.
\end{abstract}

\maketitle
Magnetic skyrmions are particle-like states~\cite{Bogdanov1994} which can occur in magnetic systems with broken inversion symmetry due to the Dzyaloshinskii-Moriya interaction~\cite{Dzyaloshinskii1958a,Moriya1960a}. Skyrmions can either form lattices at certain field ranges~\cite{Muhlbauer2009,Yu2010,Heinze2011,Yu12011,Seki2012,Romming2013,Moreau-Luchaire2016} or they can be created and manipulated individually~\cite{Romming2013}, thereby offering great potential for data storage, transfer and processing~\cite{Kiselev2011,Fert2013}. Since the geometric layout is an essential part of a device, theoretical investigations explored the effect of boundaries and confinement in skyrmionic systems~\cite{Keesman2015,Zhang2015}. For instance, skyrmion movement in racetracks~\cite{Fert2013} and the effect of notches~\cite{Iwasaki2013} on their trajectories were studied and lateral confinement was suggested as a means to stabilize skyrmions without an external magnetic field~\cite{Sampaio2013}. However,  the relative orientation of skyrmion lattices in regard to the  geometry of the boundary of magnetic systems is still an open question. Most experiments focused on extended films and on their temperature and field-dependence while finite size effects were explored only very recently, e.g.\ in nanostripes~\cite{haifeng2015} or in disk-shaped structures~\cite{beach2016,zhao2016}.

\begin{figure*}[!htb]
\includegraphics[width=\textwidth]{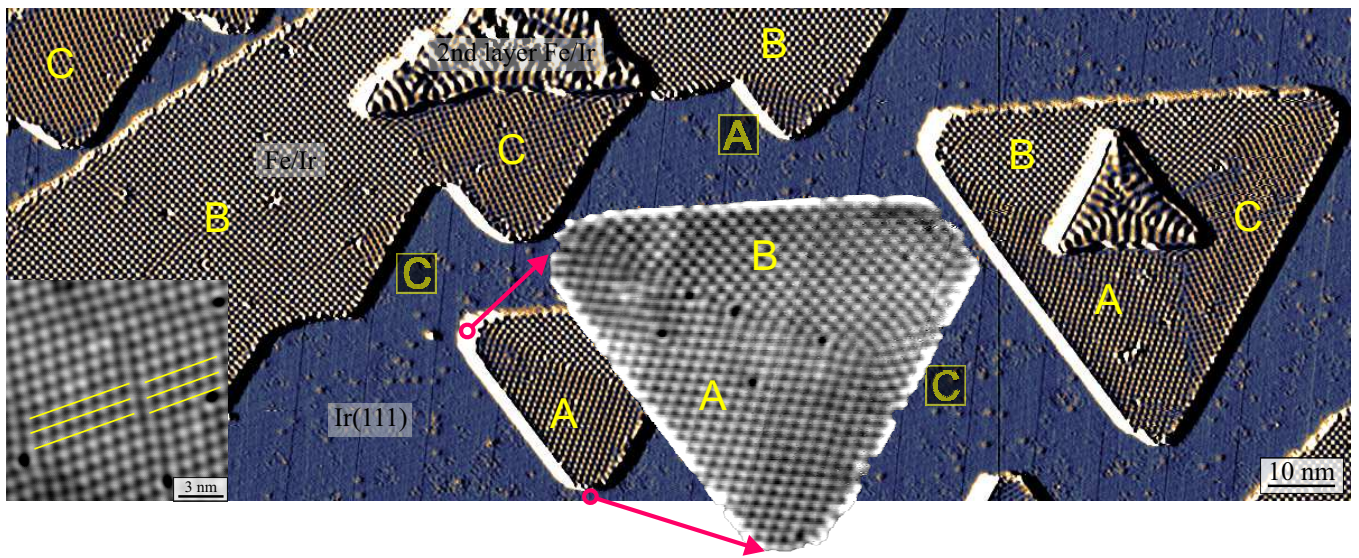}
		\caption{(color online) Coupling of nanoskyrmion lattice to close-packed edges. Spin-polarized STM current map of fcc Fe on Ir(111), the insets show topographic data in gray. Measurement parameters: $I=1$\,nA, $U=20$\,mV, $T=7.7$\,K, $B=1.5$\,T, bulk Cr tip. We observe three rotational domains of the nanoskyrmion lattice, labeled A, B and C, which show a strong correlation to the Fe step edge direction, i.e.\ one diagonal of the magnetic unit cell shows a preferrence for coupling parallel to the edge. The mismatching symmetry of the square nanoskyrmion lattice and the triangular shape of islands leads to frustration and the formation of domain walls.}
		\label{graphic:fig1}
\end{figure*}

\begin{figure}[!htb]
\includegraphics[width=\columnwidth]{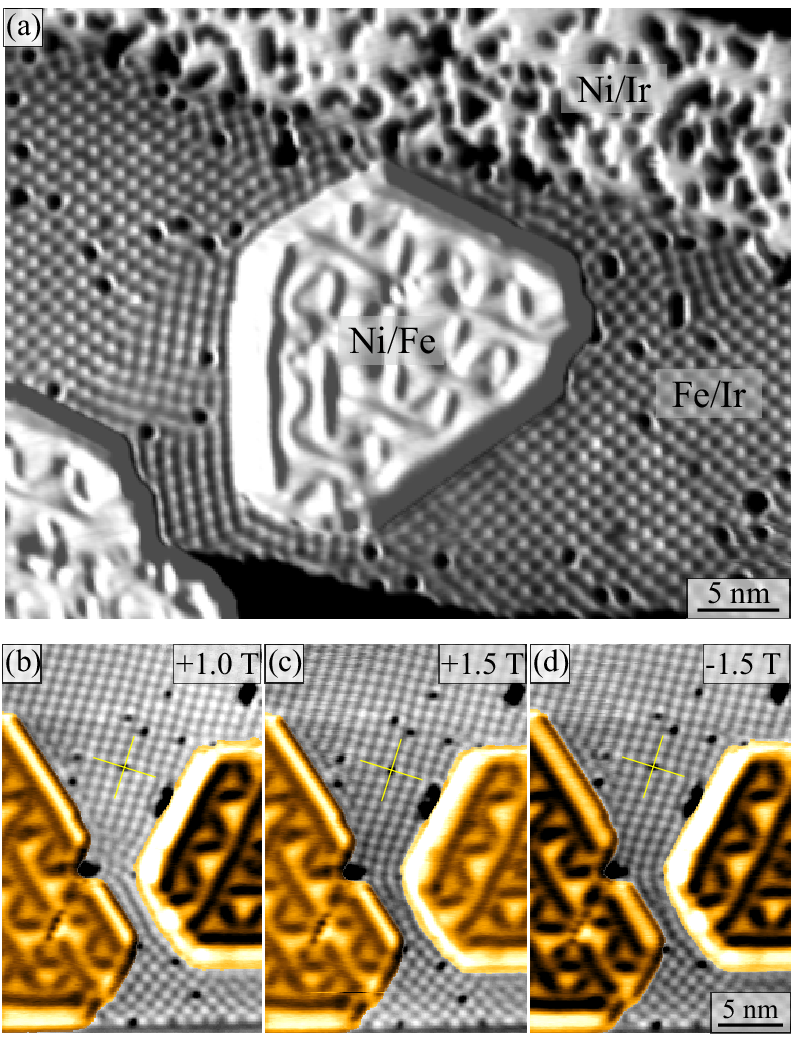}
		\caption{(color online) Coupling of nanoskyrmion lattice to ferromagnetic edge. (a) Pseudo 3D SP-STM topography image of a Ni island on a fcc Fe wire on Ir(111). The ferromagnetic Ni causes frustration in the surrounding nanoskyrmion lattice. Here, a side of the magnetic unit cell shows a preference for parallel orientation to the ferromagnetic edge, in contrast to the free-standing edges in Fig.\,1. (b-d) Field-dependent SP-STM data ($I=1$\,nA, $U=500$\,mV, $T=7.5$\,K, bulk Cr tip, gray: topographic data on Fe/Ir(111), color: d$I$/d$U$ signal on Ni/Fe/Ir(111)). (c)  The right island switched from down to up and the magnetic contrast on Fe/Ir(111) inverted in the upper part of the image. (d) Both islands switched from up to down and the magnetic contrast on Fe/Ir(111) inverted again. This can be seen by using a defect (see e.g.\ yellow cross) as a reference, and demonstrates that this part of the nanoskyrmion lattice is magnetically coupled to the right islands.}
		\label{graphic:fig2}
\end{figure}

Here, we employ spin-polarized scanning tunneling microscopy (SP-STM) and Monte-Carlo (MC) simulations to investigate the effect of two types of boundaries onto the nanoskyrmion lattice in the Fe atomic monolayer on Ir(111). This skyrmion lattice has four-fold symmetry, a period of about 1\, nm, and exists already in zero field due to the four-spin interaction~\cite{Heinze2011}. At $T=4.2$\,K it is insensitive to fields of up to 9\, Tesla~\cite{Bergmann2014}, which shows that it has a vanishing net moment. In extended films, three rotational domains can be found~\cite{Heinze2011}, where a diagonal of the magnetic unit cell aligns with one of the three symmetry equivalent close-packed atomic rows of the hexagonal Fe layer. Our new data show that at a close-packed edge of an Fe nanostructure, one diagonal of the magnetic unit cell is preferentially oriented parallel to the edge, i.e.\ the edge selects one of the three rotational domains. In triangular islands, this coupling leads to frustration and triple-domain states with domain wall widths on the order of the skyrmion size. For the more common hexa\-gonal skyrmion lattices, such a frustration effect might occur, e.g., in rectangular-shaped structures. The MC simulations with open boundary conditions already exhibit the same kind of coupling, but with unaltered interactions at the edge, multi-domains are metastable in triangular structures, i.e.\ a single-domain state has the lowest energy. This metastability arises from the spatially inhomogeneous, periodic energy distribution of the nanoskyrmion lattice, which creates additional energy barriers. The second edge we investigate is ferromagnetic with out-of-plane easy axis. Here, we find a preferred orientation of the skyrmion lattice with one side of the magnetic unit cell parallel to the edge, both in the experiment and the MC simulation. This rotational domain is neither found in extended films nor in islands, showing that controlling boundary conditions is a way to tailor specific properties in skyrmionic systems.

The Fe was evaporated onto the clean Ir(111) surface at room temperature to ensure island growth, see supplemental material for details. Fig.\,1 shows a current map of a sample area exhibiting both an extended fcc Fe atomic monolayer high wire, which has grown from an Ir step edge (upper left of Fig.\,1), and two free-standing fcc Fe islands. A current map measured with the feedback loop active, as in Fig.\,1, is essentially equivalent to a differentiated topography image, d$z$/d$x(x,y)$, and allows to show small-scale structures on large areas and on different height levels without the need for image processing. Three types of rotational domains can be seen and have been labeled  A, B and C. The domain lengths within wires can easily reach 100\,nm, see supplemental Fig.\,S1. Commonly observed one-dimensional defects of the magnetic texture within the wires are dislocation lines, see left inset, separating equivalent but laterally shifted skyrmion lattices. Apparently they result from stress, which might arise from a varying wire width or from pinning at remaining defects within the layer. The rotational domains exhibit a clear trend of coupling with a diagonal of the magnetic unit cell parallel to straight, close-packed edges: in the wire in Fig.\,1, two C domains and an A domain are found at respective edges, while the surrounding domain is of type B. In the case of the C domains a defect and the 2nd layer Fe island might have played a role in the domain formation, but no defects are found in the vicinity of the A domain. This kind of edge domain is much rarer in wires with smoother edges, see supplemental Fig.\,S1.
In the islands, due to their defined shapes, the coupling of the skyrmion lattice to the edges is most apparent: the A domain is found at the left edge, type B and C at the upper and right edges, respectively. Consequently, a triple-domain state arises from the combined effect of coupling to the edges and the mismatching symmetries of triangular island shape and square skyrmion lattice. Details of this state can be seen in the inset in Fig.\,1, showing the height data, $z(x,y)$. The domain wall width is on the order of the skyrmion size, i.e.\ $\approx 1$\,nm. The C domain is much smaller than the other two domains, possibly because the domain wall between domain A and B is pinned at defects. The magnetic frustration of the system also becomes apparent by magnetic noise within the image data, i.e.\ the spin configuration is not entirely stable during imaging, especially in the right corner of the island and at the domain wall between domain B and C. The simplest type of this kind of magnetic jitter is a domain wall movement, activated thermally or by the tunnel current. The larger island in Fig.\,1, shows a similiar triple-domain state, despite the second layer island on top, which itself has a magnetic spiral state~\cite{Hsu2016}. A further Fe island is shown in supplemental Fig.\,S2. Note that the vast majority of Fe islands grow in hcp stacking on Ir(111)~\cite{Bergmann2015}, see supplemental Fig.\,S1, and have a hexagonal spin texture, and consequently show no sign of frustration due to confinement.

The island edges select the adjacent rotational domains, despite the resulting energy cost for domain wall formation due to frustration in the interior of the island. To investigate the impact of edge properties onto the nanoskyrmion lattice, we modify the boundary condition from open to ferromagnetic. To that end, we prepared atomic layer Ni islands on fcc Fe/Ir(111) wires, which are ferromagnetic with an out-of-plane easy axis~\cite{iaia2016}. The effect of such an island onto the surrounding skyrmion lattice can be seen in Fig.\,2(a). Small domains are formed in the island's vicinity, with a side of the magnetic unit cell oriented  parallel to the island edge. These orientations are distinct from the three observed in Fe wires and islands in Fig.\,1. The coupling of the skyrmion lattice can be demonstrated directly by switching the NiFe islands in an external magnetic field: between Fig.\,2(b) and (c) the right island is switched from down to up by increasing the external field from +1\,T to +1.5\,T. This is accompanied by magnetic contrast inversion of the skyrmion lattice in the upper half of the image. Switching both islands from up to down in Fig.\,2(d) also switches the skyrmion lattice back to a state like in Fig.\,2(b). The skyrmion lattice in the upper part of the image is thus coupled to the right island, where a side of the magnetic unit cell runs parallel to the island edge.

To understand these experimental findings, we employ atomistic Monte-Carlo simulations with a single-spin Metropolis update mechanism. The magnetic monolayer Fe/Ir(111) can be described by the Hamiltonian

\begin{multline}
H = -J_\mathrm{i,j} \sum_\mathrm{i,j}\textbf{S}_\mathrm{i} \cdot \textbf{S}_\mathrm{j} -\sum_\mathrm{i,j} { \bf D}_\mathrm{i,j} \cdot \left( {\bf S}_\mathrm{i} \times { \bf S}_\mathrm{j} \right) \\ - \sum_\mathrm{ijkl} K_\mathrm{ijkl}[({\bf S}_\mathrm{i}{ \bf S}_\mathrm{j})({ \bf S}_\mathrm{k}{ \bf S}_\mathrm{l})  + ({\bf S}_\mathrm{i}{ \bf S}_\mathrm{l})({ \bf S}_\mathrm{j}{ \bf S}_\mathrm{k}) - ({\bf S}_\mathrm{i}{ \bf S}_\mathrm{k})({ \bf S}_\mathrm{j}{ \bf S}_\mathrm{l})]  \\+ \sum_\mathrm{i,j}{B_\mathrm{i,j}({\bf S}_\mathrm{i}{\bf S}_\mathrm{j})^2} + K_ \perp \sum{ (\mathrm{S}_\mathrm{i}^\mathrm{z})^2}
\end{multline}

with localized Heisenberg spins ${\bf S}_\mathrm{i}$ on a triangular lattice as demonstrated by Heinze {\em et al.}~\cite{Heinze2011}. The con\-tri\-bu\-ting energy terms originate from the exchange interaction up to the eighth nearest neighbor, the Dzyaloshinskii-Moriya (DM) interaction, the four-spin interaction, the biquadratic interaction and a perpendicular anisotropy with the parameters $J_1$ = $5.7\, \mathrm{meV}$, $J_2$ = $-0.84\, \mathrm{meV}$, $J_3$ = $-1.45\, \mathrm{meV}$, $J_4$ = $-0.06\, \mathrm{meV}$, $J_5$ = $0.2\, \mathrm{meV}$, $J_6$ = $0.2\, \mathrm{meV}$, $J_7$ = $-0.2\, \mathrm{meV}$, $J_8$ = $0.5\, \mathrm{meV}$, $D_\mathrm{i,j}$ = $-1.8\,\mathrm{meV}$, $K_\mathrm{ijkl}$ = $-1.05\,\mathrm{meV}$, $B_\mathrm{i,j}$ = $-0.2 \,\mathrm{meV}$, $K_ \perp$ = $-0.8\,\mathrm{meV}$ \cite{Heinze2011}. For simplicity, we only take contributions of the exchange interaction up to the third nearest neighbors into account. The resulting spin texture, supplemental Fig.\,S4, and the transition temperature of $T_{\rm c}=37$\,K are in good agreement with previous theoretical and experimental investigations~\cite{Heinze2011,Sonntag2014}.

\begin{figure}[!htb]
\includegraphics[width=\columnwidth]{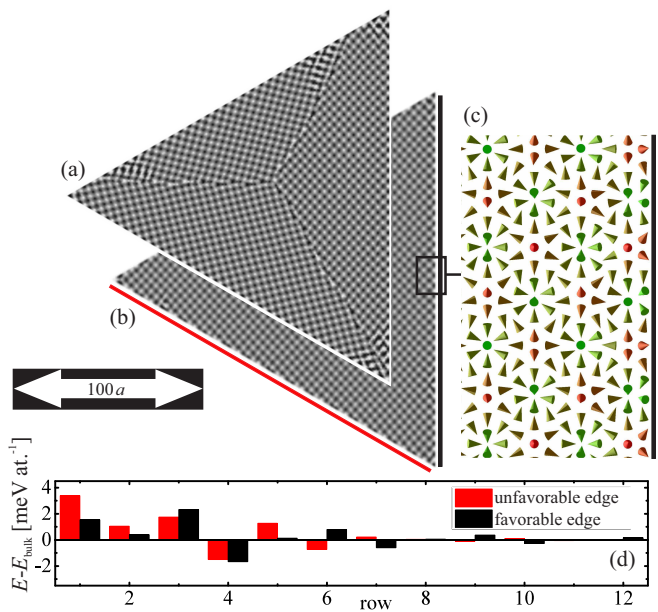}
		\caption{(color online) (a), (b) Triangular islands with open boundary conditions exhibiting multi- and single-domain states at $T = 0.1\,\mathrm{K}$. The displayed out-of-plane sensitive SP-STM images have been calculated from the Monte-Carlo spin configurations as described in Ref.\,\onlinecite{heinze2006}. (c) Spin structure of the nanoskyrmion lattice at the energetically favorable edge, taken from (b). (d) Average energy cost per atom in the n-th atomic row parallel to the edge for the two sides marked in (b) with respect to the corresponding value in the interior of a very large sample. The energetically (un)favorable edge is marked in (red) black in (b).}
		\label{graphic:fig3}
\end{figure}

\begin{figure}[!htb]
\includegraphics[width=\columnwidth]{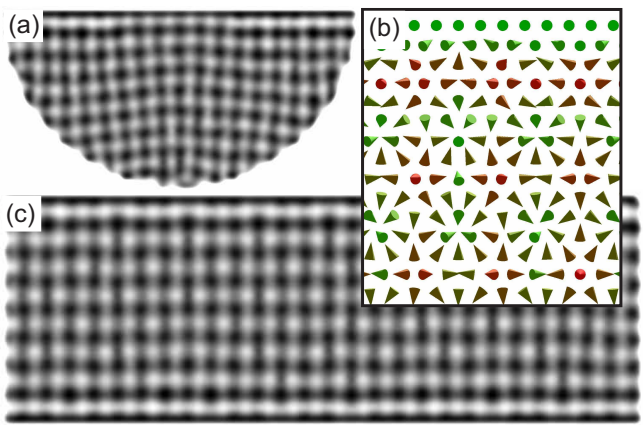}
		\caption{(color online) (a) Perpendicular magnetic contrast of the nanoskyrmion state on a half-disk with a ferromagnetic edge as obtained from MC calculations at $T = 1\,\mathrm{K}$ \cite{heinze2006}. (b) Spin structure at the upper ferromagnetic edge, taken from (c). (c)~Magnetic stripe at $T = 0.1\,\mathrm{K}$ with two ferromagnetic edges (top/bottom) and open boundary conditions left and right.}
				\label{graphic:fig4}
\end{figure}

In the experiment, we have seen a clear trend of close-packed edges selecting one of the three rotational domains, which leads to domain walls in triangular islands. To determine the energy at edges and within domain walls, and to identify the contribution of the different energy terms, we simulate magnetic islands with a triangular boundary shape, with each side consisting of 200 atoms with a nearest-neighbor distance of $a = 2.715\,\mathrm{\r{A}}$ \cite{Heinze2011}, i.e.\ a side length of 54\,nm. When reducing the temperature from $45\,{\rm K} > T_{\rm c}$ to 0.1\,K in 45 temperature steps with $5\times 10^4$ MC steps each, we typically obtain multi-domain states similar to the one shown in Fig.\,\ref{graphic:fig3}(a). 
All three rotational domains of the nanoskyrmion lattice are visible within a single island and, like in the experiment, one diagonal of the square of the magnetic unit cell is mostly aligned parallel to the edge of the island. For this island size we do not obtain a single-domain state as a result of the cooling process from a random state. For comparison, we have artificially constructed this state by periodically repeating the spin structure of one of the domains and subsequent annealing, i.e.\ by cooling from $15\,{\rm K} < T_{\rm c}$ to 0.1\,K within 15 temperature steps of $5\times 10^4$ MC steps each. The result is a single-domain state coupled to one of the edges with a diagonal of the magnetic unit cell parallel to the edge, as before, and two edges showing spin configurations with lower symmetry, see Fig.\,\ref{graphic:fig3}(b). A closer inspection of the preferred edge in Fig.\,\ref{graphic:fig3}(c) shows that the skyrmion centers are positioned in the second row from the edge. At the edge itself the spin configuration is a cycloidal spiral running along the edge, tilted about 30$^\circ$ from the perpendicular direction. This configuration is reminiscent of the coupling of spin spirals to edges in PdFe/Ir(111)~\cite{lorenz2016} and the edge tilt observed in simulations of DM dominated systems~\cite{rohart2013}.
 
We have evaluated the mean row-wise energy per atom and averaged it over several samples in order to compare the two inequivalent edges. Fig.\,\ref{graphic:fig3}(d) shows the average energy cost per atom in the n-th atomic row parallel to the edge for the two spin configurations with respect to the corresponding value in the interior of a very large sample. This graph shows that the edge affects the spin configuration further into the interior than expected from the short range interactions. The different contributions to this graph can be seen in supplemental Fig.\,S6.  When summing the values up over the atomic rows, one obtains an energy difference of $\Delta E_{\rm edge}\approx 2.2$\,meV/$a$, i.e.\ per atom at the boundary, which is equivalent to ~8.3\,meV/nm. Supplemental table I shows, that the exchange and the DM interaction, which alone would result in a spin spiral ground state, are responsible for the coupling to the preferred edge, see Fig.\,\ref{graphic:fig3}(c), while the four-spin interaction favors the other edge. 

To calculate the specific domain wall energy, we have produced several multi-domain configurations by cooling like described above, see supplemental Fig.\,S7. From the known edge energies and the respective total domain wall lengths we estimate a specific domain wall energy of $\approx 3.1$\,meV/$a$ or 11.4\,meV/nm which is on the same order as the difference in edge energies $\Delta E_{\rm edge}$. The competition of these two energy contributions favors a mono-domain state for the triangular islands. The increased energy within the wall results from all energy terms, except the four-spin and anisotropy energy, which favor domain walls.
In the simulation, multi-domain states arise due to entropy and an intrinsic pinning of domain walls by the spin texture itself. In contrast to domain walls in e.g.\ ferromagnetic samples, here, domain wall movement is suppressed as it requires a rearrangement of skyrmions, which involves an energy barrier due to their particle-like nature. This is reflected by the rather inhomogeneous energy density distributions within a wall, as can be seen in supplemental Fig.\,S8, S9. Additional effects may play a role in the experiment. Firstly, atomic defects can cause additional domain wall pinning, see Fig.\,1. Secondly, the magnetic interactions can be different at the edge compared to the interior, which would change the energy balance between single- and multi-domain state. For example, with the calculated domain wall energy and a slightly increased value of $\Delta E_{\rm edge}\approx 2.7$\,meV/$a$, single- and multi-domain state would already be energetically degenerate in a triangular island.

Next, we investigate the effect of a ferromagnetic edge onto the nanoskyrmion lattice and start with a system with the shape of a half circle in order to avoid concurrent influences of multiple edges, see Fig.\,\ref{graphic:fig4}(a). The spins at the straight edge are fixed ferromagnetically in a direction perpendicular to the system plane.  After cooling the system from a value above $T_{\rm c}$ down to $T=1$\,K, one side and not a diagonal of the square nanoskyrmion lattice is parallel to the straight edge, see, Fig.\,\ref{graphic:fig4}(a), as experimentally observed in the NiFe/Ir(111) sample. This domain is inequivalent to the previously observed rotational domains, as can be seen by comparing Fig.\,\ref{graphic:fig4}(b) and Fig.\,\ref{graphic:fig3}(c). To estimate the energy cost of this new rotational domain, we use a larger sample with two ferromagnetic edges above and below and open boundaries left and right, see Fig.\,\ref{graphic:fig4}(c). From the enclosed nanoskyrmion lattice, we estimate an energy cost with respect to the common rotational domains of $\approx 0.25$\,meV/atom. Again, the experimental system is more complex due to the pre\-sence of multiple ferromagnetic and free edges, however, the energy cost for this new domain is reflected by the fact that not all ferromagnetic boundaries induce a reorientation, see supplemental Fig.\,S3. In a pure Fe/Ir(111) system without ferromagnetic boundaries, the increased energy of this new type of domain ensures the stability of the common rotational domains against rotation.

In conclusion, we have shown that the orientation of a skyrmion lattice can be controlled by tailoring edge properties, and that the domain formation in triangular islands is governed by mismatching symmetries of island shape and skyrmion lattice. Despite the lower energy of a single-domain state, multi-domain states arise by the combined effect of entropy and an intrinsic domain wall pinning, which results from the skyrmionic character of the spin texture. 

Financial support from the DFG in the framework of SFB668 as well as from the European Union (FET-Open MAGicSky No. 665095) is gratefully acknowledged.

\end{document}